\documentstyle[emulateapj]{article}
\input epsf

\def\spose#1{\hbox to 0pt{#1\hss}}
\def\simlt{\mathrel{\spose{\lower 3pt\hbox{$\mathchar"218$}}
     \raise 2.0pt\hbox{$\mathchar"13C$}}}
\def\simgt{\mathrel{\spose{\lower 3pt\hbox{$\mathchar"218$}}
     \raise 2.0pt\hbox{$\mathchar"13E$}}}
\def\beq{\begin{equation}}
\def\eeq{\end{equation}}
\def\bce{\begin{center}}
\def\ece{\end{center}}
\def\bea{\begin{eqnarray}}
\def\eea{\end{eqnarray}}
\def\ben{\begin{enumerate}}
\def\een{\end{enumerate}}

\def\brr{\begin{array}}
\def\err{\end{array}}

\def\nh1{n_{\rm HI}}

\def\p1dk{P_{\rm 1D}(k)}
\def\simlt{\mathrel{\spose{\lower 3pt\hbox{$\mathchar"218$}}
     \raise 2.0pt\hbox{$\mathchar"13C$}}}

\newbox\grsign \setbox\grsign=\hbox{$>$} \newdimen\grdimen \grdimen=\ht\grsign
\newbox\simlessbox \newbox\simgreatbox
\setbox\simgreatbox=\hbox{\raise.5ex\hbox{$>$}\llap
     {\lower.5ex\hbox{$\sim$}}}\ht1=\grdimen\dp1=0pt
\setbox\simlessbox=\hbox{\raise.5ex\hbox{$<$}\llap
     {\lower.5ex\hbox{$\sim$}}}\ht2=\grdimen\dp2=0pt

\begin{document} 

\title{Large scale structures in the Early SDSS:  \\
Comparison of the North and South Galactic strips}

\author{Enrique Gazta\~naga$^{a,b}$}
\affil{$^a$ INAOE, Astrofisica, Tonantzintla, Apdo Postal 216 y 51, 
 Puebla 7200, Mexico}
\affil{$^b$IEEC/CSIC,
Edf. Nexus-104-c/Gran Capita 2-4, 08034 Barcelona, Spain}



\begin{abstract}
  
  We compare the large scale galaxy clustering between the North and South
  SDSS early data release (EDR) and also with the clustering in the APM Galaxy
  Survey.  The three samples are independent and cover an area of 150, 230 and
  4300 square degrees respectively. We combine SDSS data in different ways to
  approach the APM selection. Given the good photometric calibration of the
  SDSS data and the very good match of its North and South number counts, we
  combine them in a single sample.  The joint clustering is compared with
  equivalent subsamples in the APM. The final sampling errors are small enough
  to provide an independent test for some of the results in the APM. We find
  evidence for an inflection in the shape of the 2-point function in the SDSS
  which is very similar to what is found in the APM. This feature has been
  interpreted as evidence for non-linear gravitational growth. By studying
  higher order correlations, we can also confirm good agreement with the
  hypothesis of Gaussian initial conditions (and small biasing) for the
  structure traced by the large scale SDSS galaxy distribution.

\end{abstract}

\keywords{galaxies: clustering, large-scale structure of universe, cosmology}

\section{Introduction}

The SDSS collaboration have recently made an early data release (EDR) publicly
available. The EDR contains around a million galaxies distributed within a narrow
strip of 2.5 degrees across the equator. As the strip crosses the galactic
plane, the data is divided into two separate sets in the North and South
Galactic caps. The SDSS collaboration has presented a series of analysis
(Zehavi etal 2002, Scranton etal 2002, Connolly etal 2002, 
Dodelson etal 2002, Tegmark etal 2002, Szalay etal 2002) of
large scale angular clustering on the North Galactic strip, which
contains data with the best seeing conditions in the EDR.  
Gazta\~naga (2001, hereafter Ga01) 
presented a study of  bright  ($g' \simeq 20$) SDSS galaxies in 
the South Galactic EDR strip,  centering the analysis on the
comparison of clustering to the APM Galaxy Survey (Maddox etal 1990).

In this paper we want to compare and combine the bright ($r' \simeq 19$ or $g'
\simeq 20$) galaxies in North and South strips to make a detailed 
comparison between North and South and also to the APM. 
Do the North and South strips have similar
clustering? How do they compare to previous analyses?  What does the EDR tell
us about structure formation in the Universe?  Answering these questions will
help us understanding the SDSS EDR data and, at the same time, will give us
the opportunity to test how reliable are conclusions drawn from 
previous galaxy surveys. 
In particular regarding the shape of the 2-point function
(Maddox etal 1990, Gazta\~naga \& Juszkiewicz 2001) and higher order
correlations (eg Bernardeau et al. 2002, and references therein).

This paper is organized as follows. In section \S2 we present the samples
used and the galaxy selection and number counts. Section \S3 shows the
comparison of the 2 and 3-point correlation functions. We end
with some discussion and a listing of conclusions.

\begin{figure*}
\centering{
{\epsfxsize=15cm \epsfbox{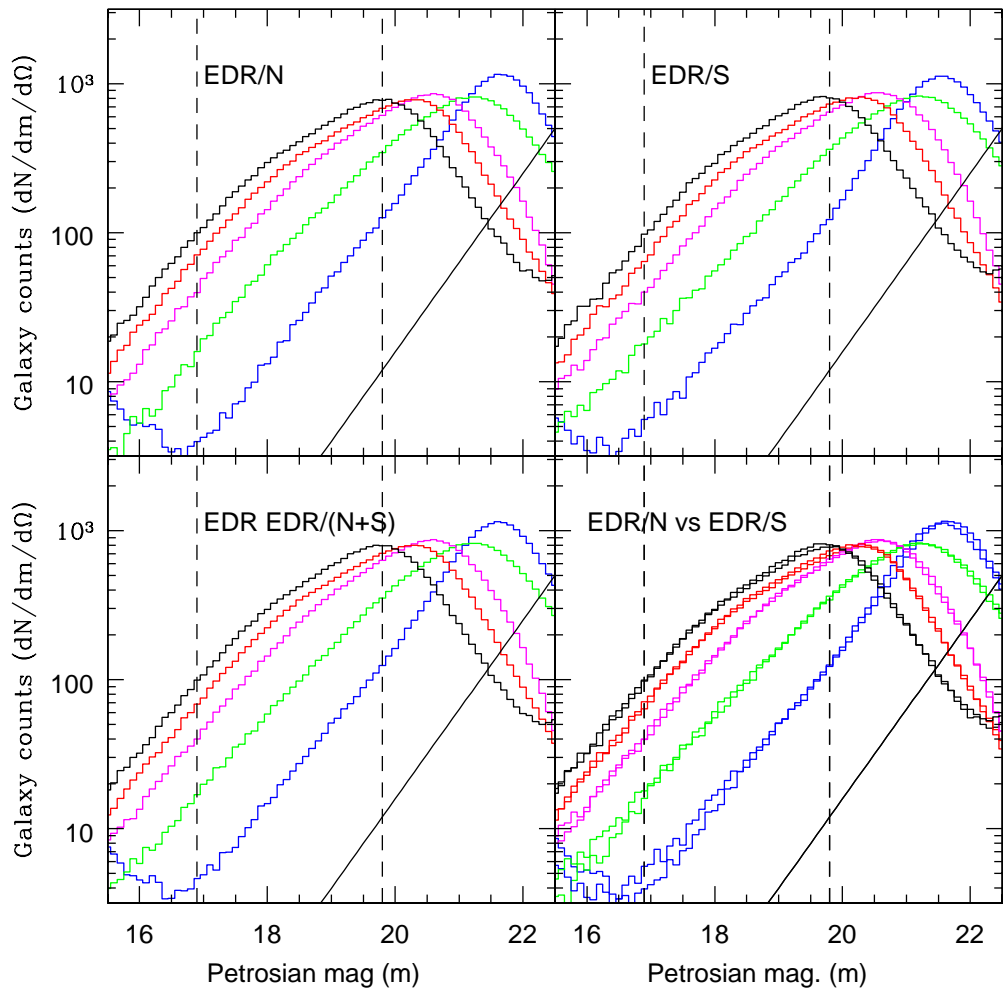}}}
\caption[Fig2]{\label{ncountsall} Galaxy density counts per 
magnitude bin and square deg. $dN/dm/d\Omega$ as a function of
Petrosian magnitude $z',i',r',g',u'$ (from left to right). The top
panels show EDR/N (left) and EDR/S (right). The bottom left panel
shows EDR/(N+S) while the right panel compares EDR/S to EDR/N.}
\end{figure*}

\begin{figure*} 
\centering{
{\epsfxsize=18cm \epsfbox{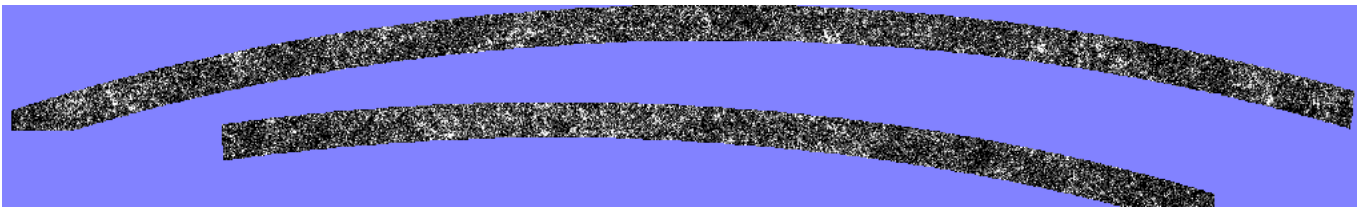}}
}
\caption[Fig2]{\label{MapNSGCg20c} 
Pixel maps of equatorial projections 
of EDR/N (top) with
 $2.5 \times 90$ sqr.deg. and EDR/S (bottom)
 with $2.5 \times 60$ sqr.deg.}
\end{figure*}

\section{SDSS samples and pixel maps}

We follow the steps described in Gazta\~naga (2001, hereafter Ga01).  We
download data from the SDSS public archives using the SDSS Science Archive
Query Tool (sdssQT, http://archive.stsci.edu/sdss/software/).  We select
objects from an equatorial SGC (South Galactic Cap) strip 2.5 wide ($-1.25 <
DEC <1.25$ degrees.)  and 66 deg. long ($351 < RA < 56 $ deg.), which will be
called EDR/S, and also from a similar NGC (North Galactic Cap) 2.5 wide and 91
deg.  long ($145 < RA < 236 $ deg.), which will be called EDR/N.  These strips
(SDSS numbers 82N/82S and 10N/10S) correspond to some of the first runs of the
early commissioning data (runs 94/125 and 752/756) and have variable seeing
conditions.  Runs 752 and 125 are the worst with regions where the seeing
fluctuates above 2''.  Runs 756 and 94 are better, but still have seeing
fluctuations of a few tenths of arc-second within scales of a few
degrees\footnote{See http://www-sdss.fnal.gov:8000/skent/seeingStatus.html or
  Figure 4 in Scranton etal 2001}.  These seeing conditions could introduce
large scale gradients because of the corresponding variations in the
photometric reduction (eg star-galaxy separation) that could manifest as large
scale number density gradients (see Scranton et al 2001 for a detailed account
of these effects).  We will test our results against the possible effects of
seeing variations, by restricting the analysis to runs 756 and 94, and by
using a seeing mask (see \S 3.3).

We will also consider a sample which includes both the North and South strips,
this will be called: EDR/(N+S). Note that the clustering from this sample will
not necessarily agree with the mean of EDR/N and EDR/S, eg EDR/(N+S) $\ne$
EDR/N+EDR/S (see below).

We first select all galaxies brighter than $u'=22.3, g'=23.3, r'=23.1,
i'=22.3, z'=20.8$, which corresponds to the SDSS limiting magnitudes for 5
sigma detection in point sources (York etal. 2000). Galaxies are found from
either the $1\times 1$, $2\times 2$ or $4\times 4$ binned CCD pixels and they
are de-blended by the SDSS pipeline (Lupton etal 2001).  Only isolated
objects, child objects  (resulting from deblending)
and objects on which the de-blender gave up are used in
constructing our galaxy catalog
(see Yasuda etal 2001). 

There are about 375000 objects in our sample
classified as galaxies in the EDR/S and about
504000 in the EDR/N.  Figure \ref{ncountsall} shows the number counts (surface
density) for all these 879000 galaxies as a function of the magnitude in each
band, measured by the SDSS modified Petrosian magnitudes $m_u'$, $m_g'$,
$m_r'$, $m_i'$ and $m_z'$ (see Yasuda et al 2001 for a discussion of the SDSS
counts).  Continuous diagonal lines show the $10^{0.6 m}$ expected for a low
redshift homogeneous distribution with no k-correction, no evolution and
no-extinction.

We next select galaxies with SDSS modified Petrosian magnitudes to match the
APM selection: $17<B_J<20$, which corresponds to a mean depth of ${\cal D} \sim
400$ Mpc/h. We try different prescriptions.  We first apply the following
transformation to mimic the APM filter $B_J$:

\beq
B_J = g' + 0.193 (g'-r') + 0.115
\label{BJ}
\eeq

This results from combining the relation $B_J =B - 0.28 (B-V)$ (Maddox etal
1990) with expressions (5) and (6) in Yasuda etal (2001). As the mean color
$g'-r' \simeq 0.7$ the above relation gives a mean $B_J \simeq g' +0.25$,
which roughly agrees with the magnitude shift used in Ga01. For the $17<B_J<20$
range (using the above transformation) we find $N\simeq 123000$ galaxies in
the EDR/(N+S), with a galaxy surface which is very similar to the one in the
APM (only $5\%$ larger after substraction of the $5\%$ star-merger
contribution in the APM).  In any case, this type of color transformations
between bands are not accurate and they only work in some average statistical
sense.  The uncertainties are even larger when we recall that the APM uses
fix isophotal aperture, while SDSS is using Petrosian magnitudes, a difference
that can introduce additional color terms and surface brightness dependence.

It is much cleaner to use a single SDSS band. We should use  $g'$ which
is the closest to the APM $\lambda_{B_J}\simeq 4200 A$ ($\lambda_{u'}\simeq
3560$, $\lambda_{g'}\simeq 4680$ and $\lambda_{r'}\simeq 6180$). But how do we
decide the range of $g'$ to match the APM $17<B_J<20$? We try two
approaches.  One is to look for the magnitude interval that has the same counts, as
done in Ga01. The resulting range is $16.8<g'<19.8$. This gives a reasonable
match to the clustering amplitudes in the EDR/S and EDR/N.  But there is no
reason to expect a perfect match: the selection function and resulting depth
is different for different colors.  The other approach is to fix 
the magnitude range, ie $17<g'<20$, rather than the  counts. 
This produces $N \simeq 157000$ galaxies, which corresponds $\simeq 25\%$
higher counts than the APM.  This does not necessarily mean that this sample
is deeper than the APM, because of the intrinsic different in color selection,
K-corrections and possible color evolution.

Finally, we produce equal area projection pixel maps of various resolutions
similar to those made in Ga01. Except for a few tests, all the analyses
presented here correspond to 3'.5 resolution pixels.  On making the pixel maps
we mask out about 1.'75 of the EDR 
sample from the edges, which makes an integer
number of pixels in our equatorial projection.  This also avoids  potential
problem of the galaxy photometry on the edges (although higher resolution maps
show very similar results, indicating that this is not really a problem).

\subsection{Galactic extinction}

The above discussion ignored Galactic extinction. It should be noted that the
standard extinction law $A_b= C(\csc b-1)$ with $C=0.1$ was used for the APM
photometry. This is a very small correction: $A_b =0$ at the poles ($b=90$
deg) and the maximum $A_b\simeq 0.03$ at the lowest galactic declination ($b
\simeq 50$ deg). This is in contrast to the Schlegel etal (1998) extinction
maps which have significant differential extinction $E(B-V) \simeq 0.02-0.03$
even at the poles.  The corresponding total absorption $A_b$ for the $B_J$
band according to Table 6 in Schlegel etal (1998) is four times larger: $A_b
\simeq 0.08-0.12$. This increases up to $A_b \simeq 0.2-0.3$ at galactic
declination $b \simeq 50$. Thus, using the Schlegel etal (1998) extinction
correction has a large impact in the number counts for a fix magnitude
range. The change can be roughly accounted for by
shifting the mean magnitude ranges by the mean extinction, eg
$\simeq 0.2$ magnitudes in  $B_J$. It is therefore
important to know what extinction correction has been applied when comparing
different surveys or magnitude bands.

Despite the possible impact on the quoted magnitudes (and therefore counts),
extinction has little impact on clustering, at least for $r'<21$ (see Scranton
etal 2001 and also Tegmark etal 1998).  This is fortunate because of the
uncertainties involved in making the extinction maps and its calibration.
Moreover, the Schlegel etal (1998) extinction map only has a 6'.1 FWHM, which
is much larger than the individual galaxies we are interested on. Many dusty
regions have filamentary structure (with a fractal pattern) and large
fluctuations in extinction from point to point. One would expect similar
fluctuations on smaller (galaxy size) scales, which introduces further
uncertainties to individual corrections.

Here we decided as default not to correct for extinction, because this will be closer to the
APM analysis and makes little effect on clustering at the depths and for the
issues that will be explored here. This has been extensively checked for EDR/N
by Scraton et al. (2001). We have also check this here in all
EDR/N, EDR/S and EDR/(N+S), see \S\ref{sec:seeing}

To avoid confusion with other prescriptions by the SDSS collaboration
we will use $z',i',r',g',u'$ for 'raw', uncorrected magnitudes, and
$z^*,i^*,r^*,g^*,u^*$ for extinction corrected magnitudes.  For example,
according to Schlegel etal (1998) $r'=18$ corresponds roughly to an average
extinction corrected $r^* \simeq 17.9$ for a mean differential extinction
$E(B-V) \simeq 0.03$.

\begin{figure*} 
\centerline{
{\epsfxsize=9.cm \epsfbox{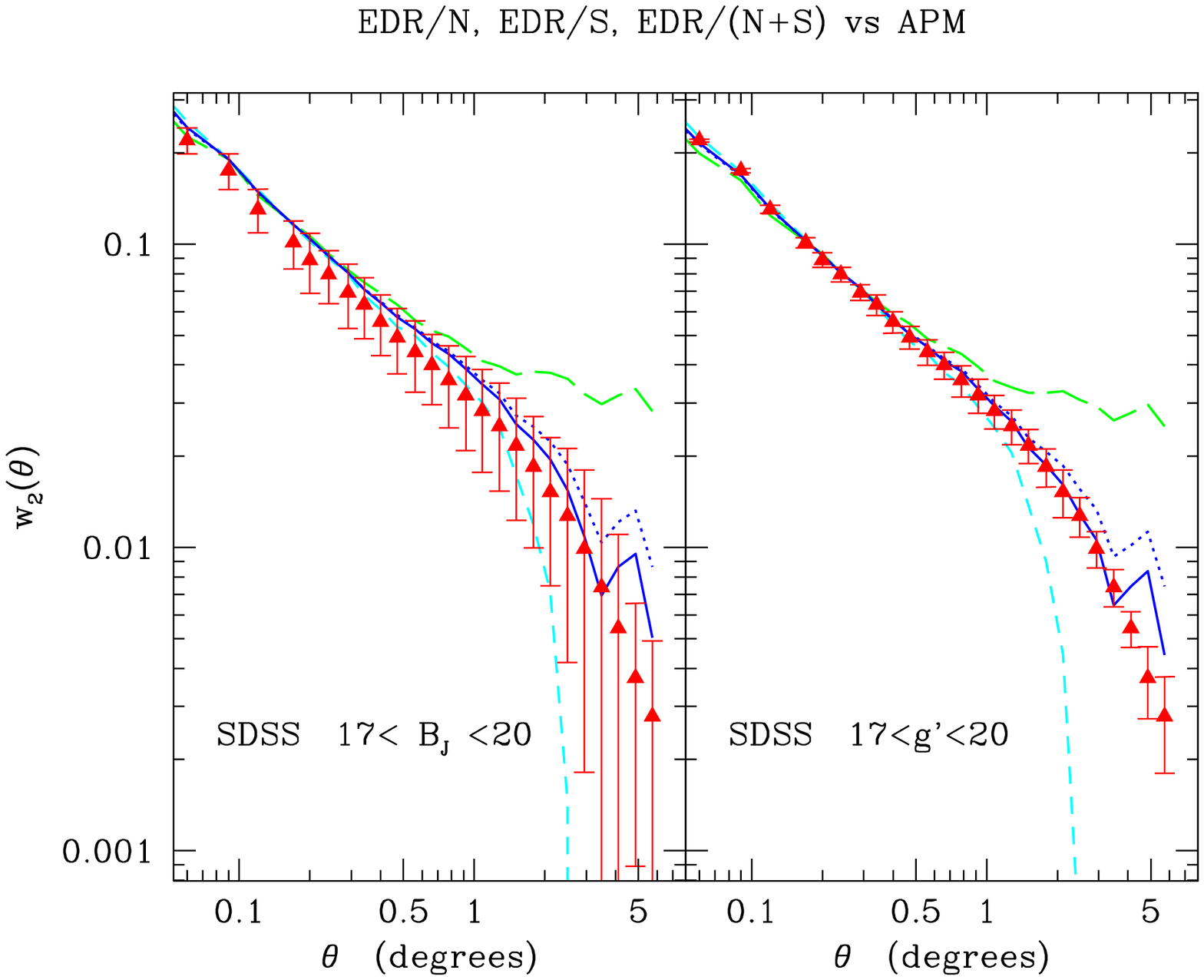}}
{\epsfxsize=9.cm \epsfbox{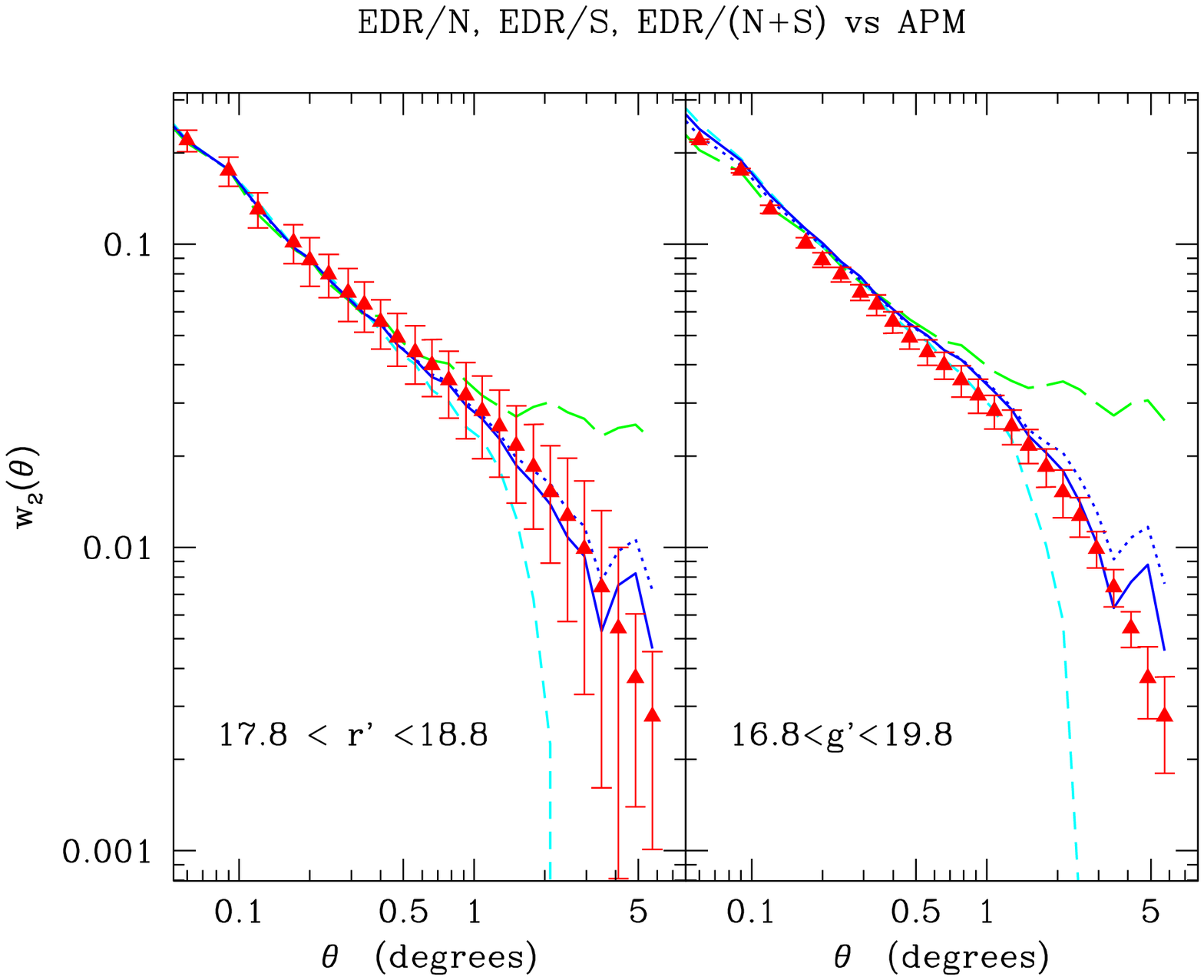}}
}
\caption[Fig2]{\label{w2sdssNS} 
The angular 2-point function  $w_2(\theta)$ as a function
of galaxy separation  $\theta$ for different SDSS magnitudes as
labeled in the figures. The short and long dashed
lines correspond to the SDSS EDR/N (North) and  EDR/S (South) strips. 
The continuous line corresponds to EDR/(N+S), a joint analysis of the
SDSS South and North data. The dotted line is the mean of the
EDR/N and EDR/S. The triangles with errorbars show
the mean and 1-sigma confidence level in the values of
10 APM sub-samples that simulate the different EDR samples:  
EDR/S (errors in the first panel),  
EDR/(N+S) (error in second and fourth panels) and EDR/S (errors in
third panel). }
\end{figure*}

\section{Clustering comparison}

To study sampling and estimation biasing effects on the SDSS clustering estimators we
have cut different SDSS-like strips out of the APM map (see Ga01). For the APM,
we have considered a $17<B_J<20$ magnitude slice in an equal-area projection
pixel map with a resolution of $3.5$ arc-min, that covers over $4300$ $deg^2$
around the SGC.  The APM sample can fit about 25 strips similar to EDR/S and
16 similar to EDR/N.  The APM can not cover the combined EDR/(N+S) as it
extends across the whole equatorial circle. But we can select several
subsamples consisting of sets of 2 strips, one like EDR/S and another one like
EDR/N well separated within the APM map, eg by at least $10$ degrees. As
correlations are negligible on angular scales $>10$ degrees, this simulates
well the combined EDR/(N+S) analysis.  To study sampling effects over
individual scans we also extract individual SDSS-like CCD scans out of the APM
pixel maps.  In all cases we correct the clustering in the APM maps for a
$5\%$ contamination of randomly merged stars (see Maddox etal 1990), ie we
scale fluctuations up by $5\%$ (see also Gazta\~naga 1994).

\subsection{The angular 2-point function}

We first study the angular two-point function.  Figure \ref{w2sdssNS} shows
the results from the EDR/S (long-dashed), EDR/N (short-dash) and EDR/(N+S)
(continuous lines). As mentioned above, the clustering from the combined
sample EDR/(N+S) will not necessarily agree with the mean of EDR/N and EDR/S
(shown as dotted lines) for several reasons: estimators are not linear,
neither are sampling errors and local galaxy fluctuations are estimated around
the combined mean density (rather than the mean density in each subsample). As
shown in Figure \ref{w2sdssNS} the two estimators yield different results. In
general for a well calibrated survey the whole, ie EDR/(N+S), should give
better results than the sum of the parts, so that we take the EDR(N+S) results
as our best estimate.

In general, the results for the $w_2$ shape in EDR/N in Figure \ref{w2sdssNS}
agree well with the corresponding comparison in Fig.1 of Connolly etal (2002),
with a sharp break to zero around 2-3 degrees.  The results for EDR/S agree
well with Ga01, showing a flattening at similar scales.  Note how EDR/(N+S)
$17<B_J<20$ (shown in the first panel) are about $15\%$ higher in amplitude
that the APM (this is not a very significant discrepancy for the EDR/S errors
shown in the plot, but it is when compared to the EDR/(N+S) errors from the
APM, shown in panels 2 and 3).  As mentioned above this is not totally
surprising as the magnitude conversion in Eq.[\ref{BJ}] could only works on
some average sense.  Results for $16.8<g'<19.8$ are intermediate between $B_J$
and $17<g'<20$.

\begin{figure*} 
\centering{
\centering{\epsfxsize=18cm \epsfbox{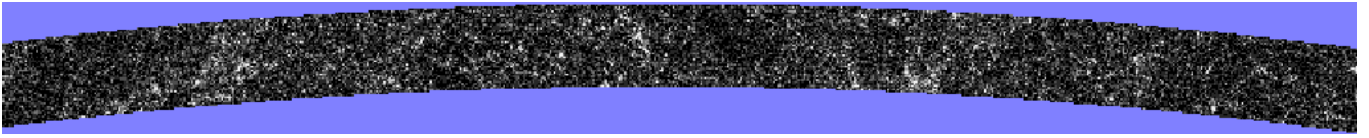}}}
\caption[Fig2]{\label{pixelmap} 
Pixel maps of the central  ($2.5 \times 40$ sqr.deg.) region of
the EDR/N strip (10N/10S) with the full 752+756 overlapping runs.} 
\end{figure*}
\begin{figure*} 
\centering{
\centering{\epsfxsize=18cm \epsfbox{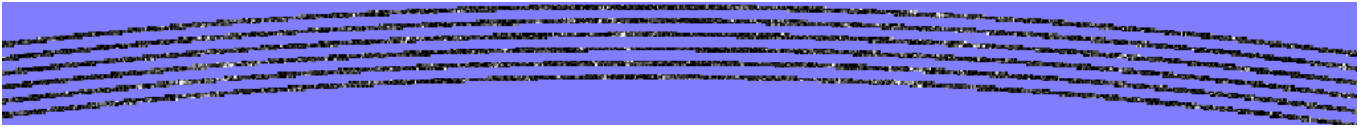}}}
\caption[Fig2]{\label{pixelmap2} 
Same as Fig.\ref{pixelmap}
but with only the central part of 6 CCD regions in
 run 756. With this resolution (3'.5)
a CCD field is only a few pixels wide.}
\end{figure*}

\begin{figure*} 
\centerline{
{\epsfxsize=9.cm \epsfbox{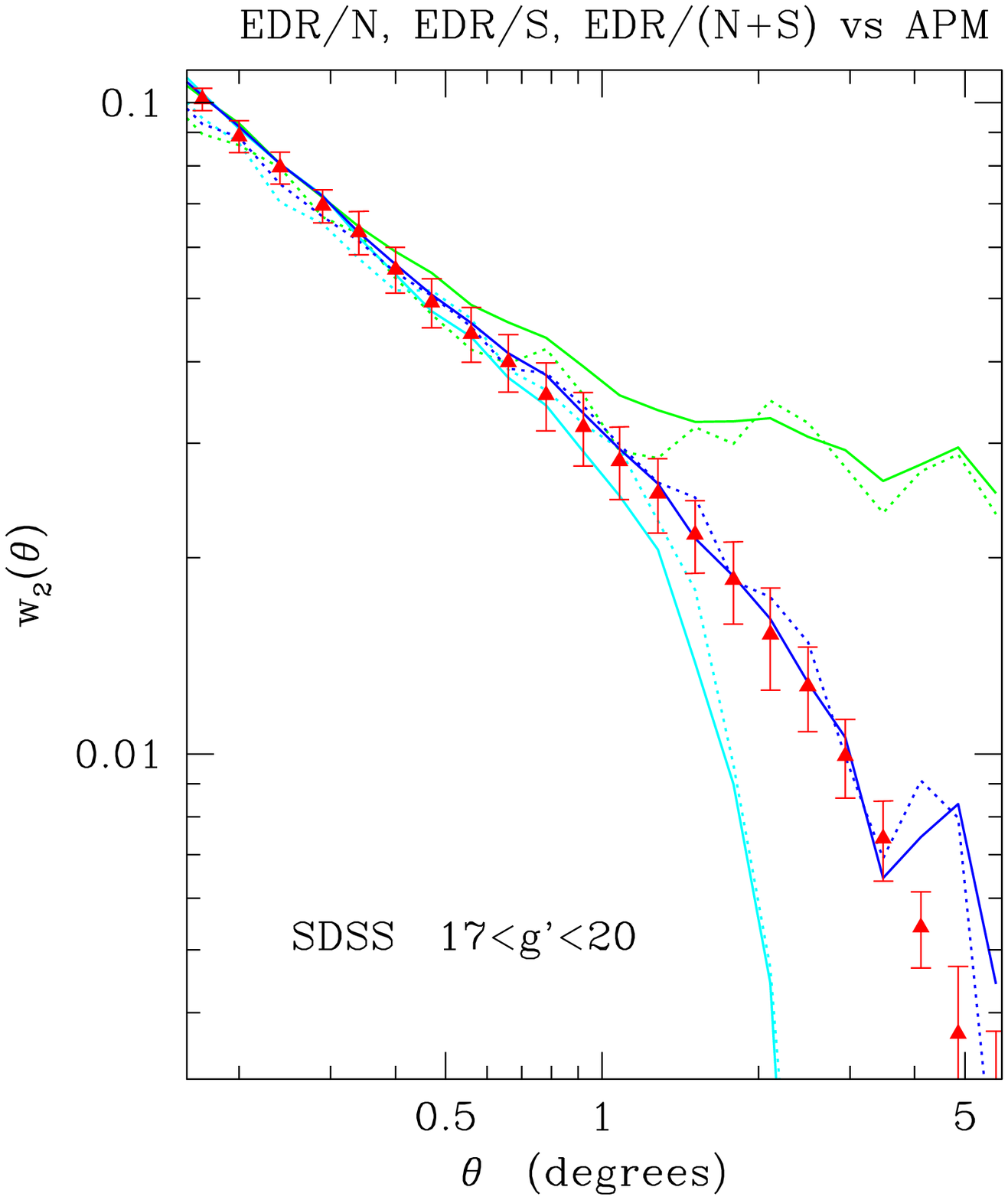}}
{\epsfxsize=9.cm \epsfbox{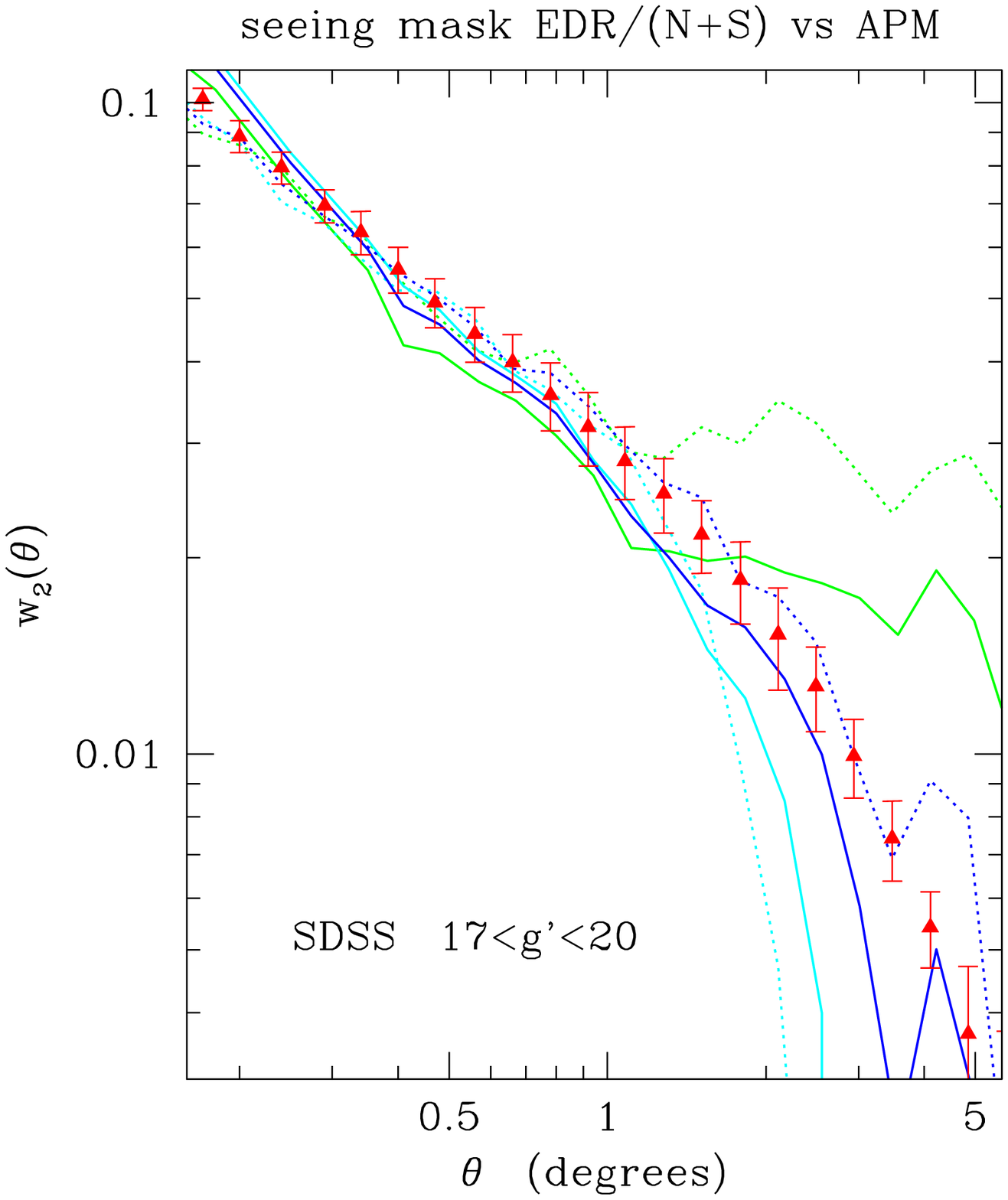}}}
\caption[Fig2]{\label{w2sdssNSscan}
  {\sc Left panel:} Zoom over a region of the second panel of Figure
  \ref{w2sdssNS}.  Here we compare the full strips of EDR/S, EDR/N and
  EDR/(N+S) (continuous lines from top to bottom at the largest angles) to the
  APM subsamples (triangle with errorbars) of size similar to EDR/(N+S). The
  dotted lines correspond to the central part of the CCD in scans 94 (top
  dotted line, next to EDR/S), 756 (bottom, next to EDR/N) and
  the joint 756+94 (middle dotted line, along EDR/(N+S)).\\
  ~\\
  {\sc Right panel:} The dotted lines are as in the left panel, while
the continuous lines correspond to
the seeing mask in Fig.\ref{MapNSGCg20sei200}.}
\end{figure*}
\begin{figure*} 
\centering{
{\epsfxsize=18cm \epsfbox{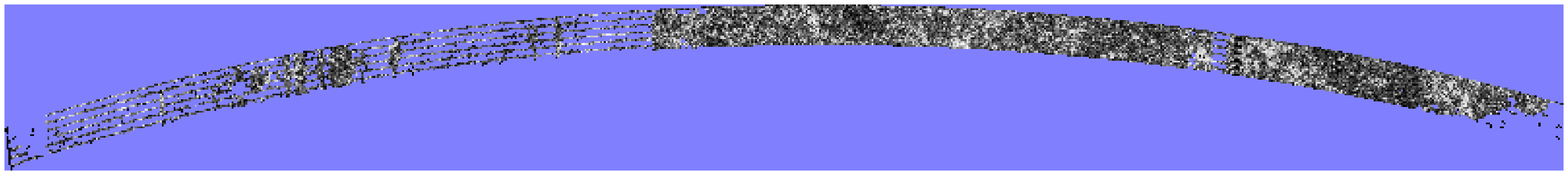}\vspace{-.5cm}}
{\epsfxsize=12cm \epsfbox{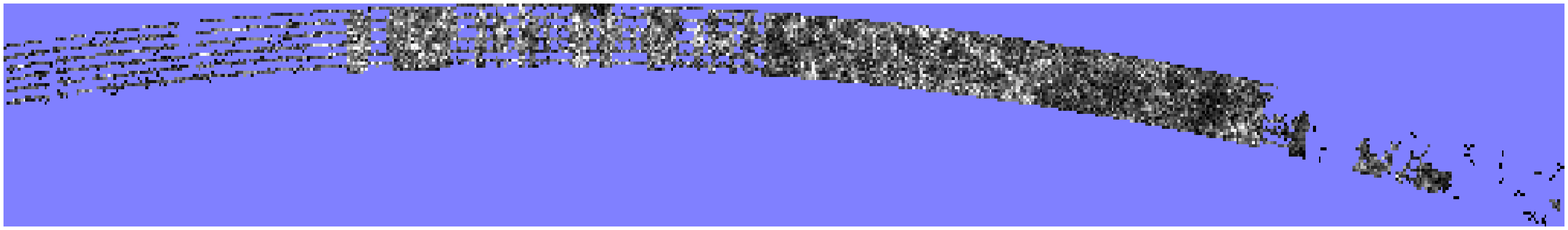}}
}
\caption[Fig2]{\label{MapNSGCg20sei200} 
Pixel maps similar to Fig.\ref{MapNSGCg20c}
with the seeing and extinction mask.} 
\end{figure*}

Scranton et al 2002 studied the SDSS systematic effects with $r^*$ colors, and
found that systematic effects had negligible contributions to $w_2(\theta)$
for $r^*<21$ (eg see their figure 15). The APM has a depth
corresponding to $r^* \simeq 18.5$, which is almost 3 magnitudes brighter than
the above limit.  Nevertheless, for comparison, we also study the
$w_2(\theta)$ shape in $r'$.  The brighter sample of $r^*=18-19$ in Connolly
etal (2001) is slightly deeper that the APM, with $z\simeq 0.18$ rather than
$z\simeq 0.15$ for the APM. We find that $r'=17.8-18.8$ is the closest one
magnitude $r'$ bin in depth to the APM.  Because of the average extinction
this corresponds roughly to extinction corrected $r^*=17.65-18.65$.  This
sample has about $40\%$ fewer galaxies (per square degree) than the APM,
presumably because of the color correction and differences in the photometric
selection.  Results for this $r'$ sample (shown in the third panel of
Figure \ref{w2sdssNS}) agree quite well with the APM amplitude. Here the errors
are from APM subsamples similar to EDR/N (note
how they are slightly smaller than the errors in the APM shown in the first
panel of Figure \ref{w2sdssNS}, as expected from the smaller area of EDR/S).

Overall, we see how the shape of the 2-point function in all EDR samples
remains remarkably similar for the different magnitudes. This is despite the
fact that the mean counts change by more than $60\%$ from case to case. The
amplitude of $w_2$ changes by about $20\%$ from one SDSS sample to the other,
but the shape remains quite similar.  The best match to the APM amplitude is
for the $17<g'<20$, which will be taken for now on as our reference
sample.

\subsection{$w_2(\theta)$ in central scans 756 and 94}

As a test for systematics, we study $w_2(\theta)$ using only the
central region of the CCD in scans 756 (North EDR) and 94 (South EDR).  The
seeing during run 756 is the best in the EDR with only small fluctuations
around 1".4. Regions in the SDSS where the seeing degraded to worse than 1".5
are marked for re-observations.  As this includes most of the EDR data one may
worry that some of the results presented here could be affected by these seeing
variations . As mentioned above this has been shown not be the case, at least
for $r^*<21$ (Scranton etal 2001).

We estimate $w_2(\theta)$ using only galaxies in the central regions of the
CCD in scans 756 (the best of EDR/N) and 94 (the best of EDR/S).  Figure
\ref{pixelmap} shows a piece of this new data set for the EDR/N.  As can be
seen in the figure (bottom panel) we only consider the central part of 756 to
avoid any contamination from the CCD edges. This new data set 
contains only $30\%$ of the area (and of the galaxies) from the whole strip.

Left panel in Figure \ref{w2sdssNSscan} 
compares the results of $w_2(\theta)$ for the
individual scans against the whole strip for all EDR/S, EDR/N and EDR/(N+S).
As can be seem in this Figure, individual scans (dotted lines) agree very well
with the corresponding overall strip values. Possible
systematic errors seem quite small.
In fact, the agreement is striking after a visual comparison of the heavy
masking in the pixel maps of Figure \ref{pixelmap} (which shows the actual
resolution use for the $w_2(\theta)$ estimation in Fig. \ref{w2sdssNSscan}).
One would naively expect some more significant sampling variations when we use
only 1/3 of the data.  But nearby regions are strongly correlated and we can
get very similar results with only a fraction of the data (this is also nicely
shown in Figures 13-15 of Scranton etal 2001).  This test shows the power of
doing configuration space analysis (as opposed to Fourier space analysis,
eg see Scoccimarro etal 2001).  It
also illustrates that our estimator for $w_2(\theta)$ performs very well on
dealing with masked data.

\subsection{Seeing and reddening mask}
\label{sec:seeing}

Fig.\ref{MapNSGCg20sei200} shows athe pixels EDR/N and EDR/S with a seeing
better than 2 arc-sec. and 0.2 maximum extinction. Pixels with larger seeing or
larger extinction are masked out. As apparent in the Figure there is a
significant reduction of the available area after the masking.  Right panel of
Figure \ref{w2sdssNSscan} compares the results of $w_2(\theta)$ for the new
masked maps with the results for the individual scans, for all EDR/S, EDR/N
and EDR/(N+S). There is now a much better agreement between the EDR/N and
EDR/S, which suggest that the discrepancies between EDR/N and EDR/S apparent
in the left panel of Fig.\ref{MapNSGCg20sei200} are due to these systematic
effects. The number of available pixels ($30 \%$ of the total)
is comparable to the ones in
individual scans (dotted line), which indicates that samplings errors can not
account for the observed differences between the dotted and continuous lines.
The biggest change is apparent in EDR/S, which is the smallest sample
and the one subject to worst seeing conditions. 
Most of the difference is due to
the seeing rather than the extinction mask.

  We find similar results for slightly lower cuts in seeing and
extinction, but the number of pixels in EDR/S decrease very quickly as we
lower the seeing, and sampling errors dominate over any possible systematics.
Thus such a tests are not very conclusive. 

Note that the APM absolute 
errors should  be larger for the masked data, as there
are less area available. Note also that the mean EDR/(N+S) is lower 
(continuous middle line in the right panel of Figure \ref{w2sdssNSscan}).
The absolute samplings error (as oppose to the relative errors) 
approach a constant on $\theta \sim 1$ deg. scale
(e.g.  Fig.\ref{varw2} and Eq.\ref{varw2}). All of this indicates that
the final relative errors should  be larger. Thus, taking into account
 these considerations, the discrepancies between EDR/N and EDR/S are 
not significant anymore, and certainly within 2-sigma errors from the APM.

\subsection{Variance and covariance}

\begin{figure*}
\centerline{
{\epsfxsize=8cm \epsfbox{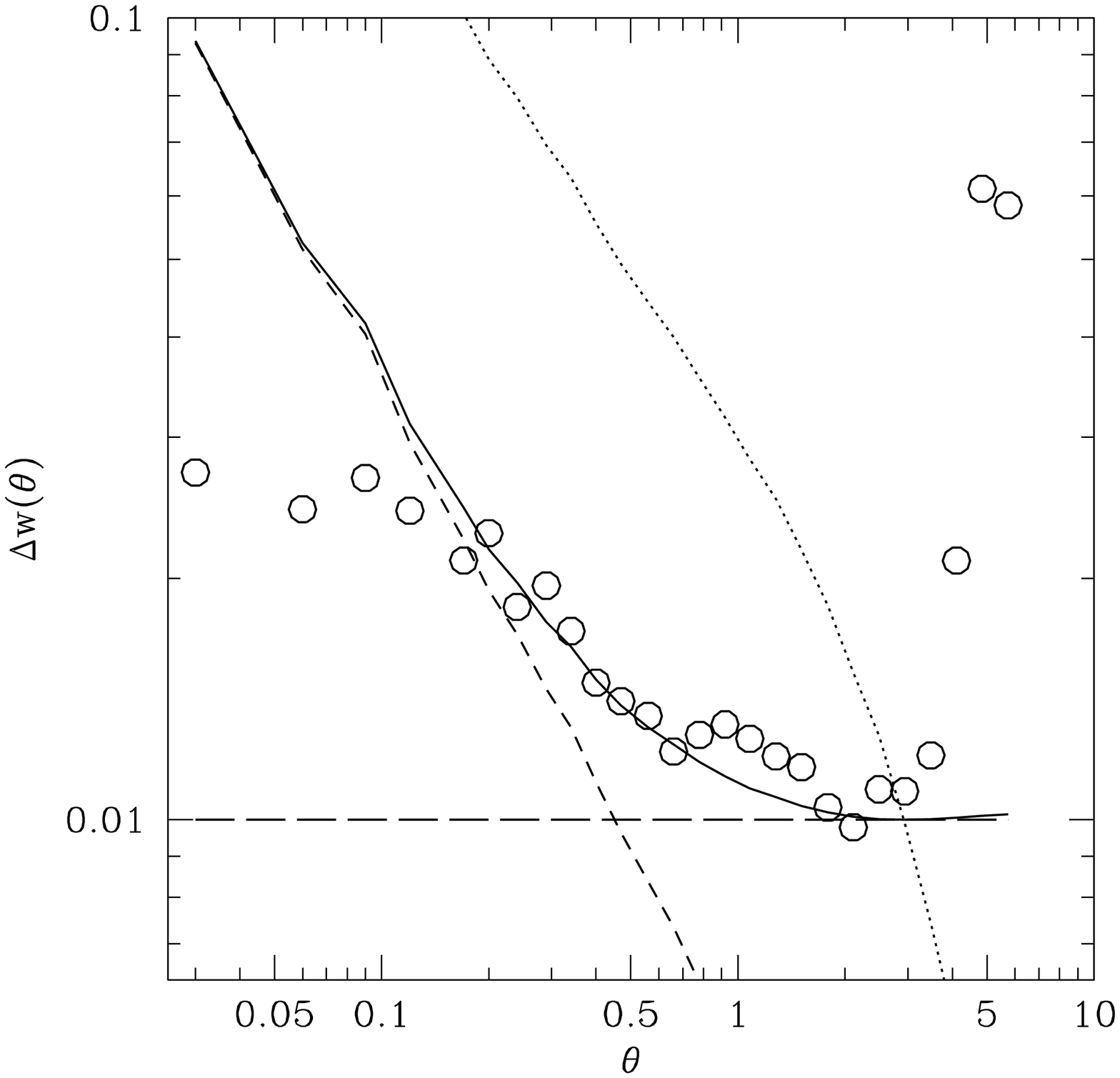}}
{\epsfxsize=8cm \epsfbox{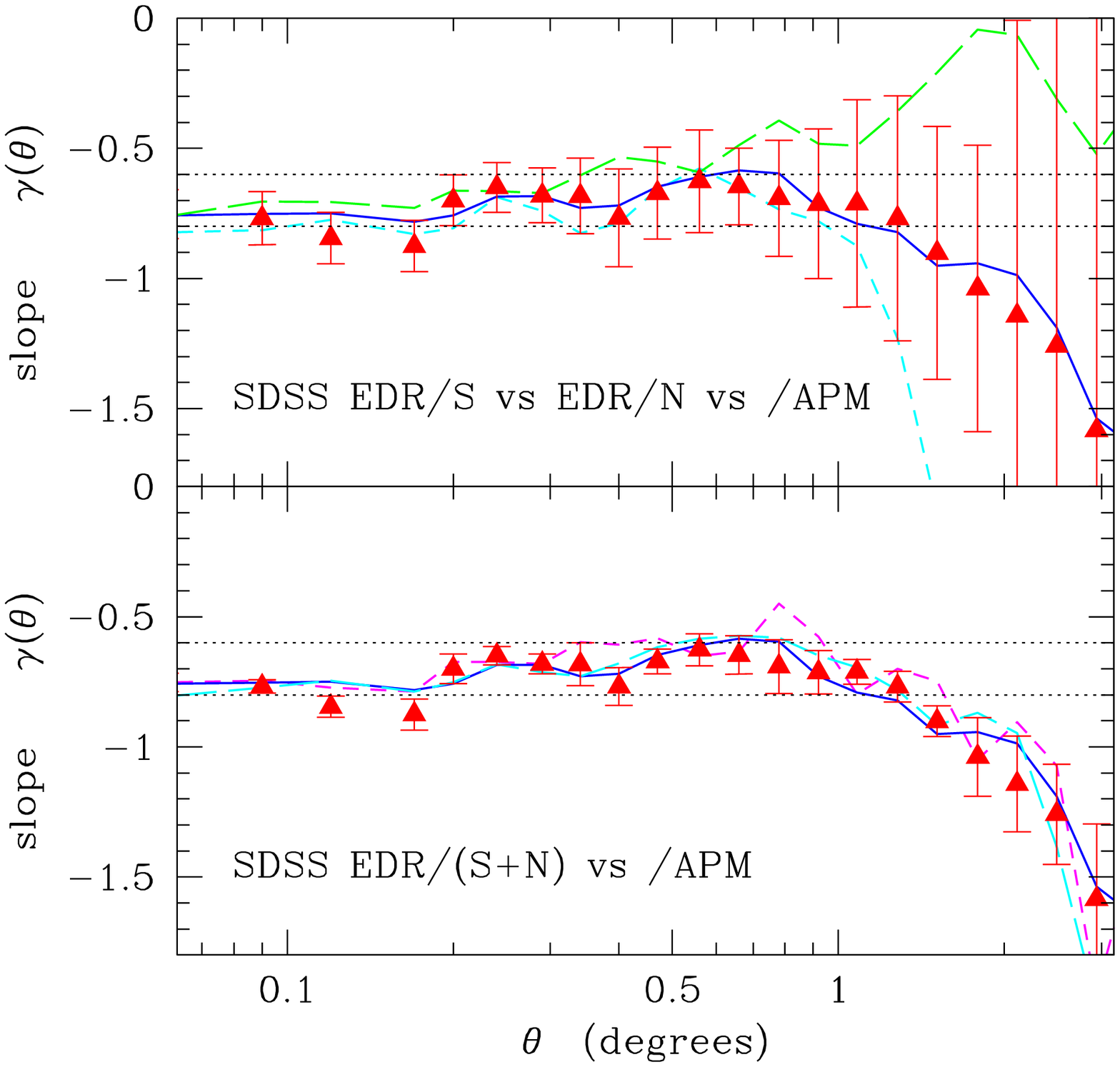}}}
\caption[Fig3]{\label{varw2} 
 {\sc Left panel:} Mean (dotted line) and
errors (circles) in $w(\theta)$ from
10 APM sub-samples that simulate EDR/N. The errors are 
compared with the theoretical expectation in Eq.\ref{var} (continuous
line). The dashed lines show the contributions from each 
of the terms in Eq.\ref{var}. 
  ~\\
  {\sc Right panel:} Logarithmic slope of the angular 2-point function
  $\gamma(\theta)$ as a function of galaxy separation $\theta$ for SDSS.  The
  lines in the top panel correspond to the ones in the second panel of
  Fig.\ref{w2sdssNS} (ie EDR/S, EDR/N and EDR/(N+S) in $17<g'<20$).  In the
  bottom panel we also show the EDR/(N+S) in the corrected $17<B_J<20$ (long
  dashed line), again $17<g'<20$ (continuous line) and the central region of
  the CCDs in scans 756+94 (short dashed line). The triangles with errorbars
  show the mean and 1-sigma confidence level in the values of several APM
  sub-samples similar to EDR/(N+S). The errorbars in the top panel corresponds
  to APM sub-samples similar to EDR/S.  The horizontal dotted lines corresponds
  $\gamma = -0.6$ and $\gamma = -0.8$.}
\end{figure*}

Bernstein (1994) has calculated the covariance in the angular 2-point function
$w_i \equiv w(\theta_i)$, where $\theta_i$ correspond to the bins in angular
separation (for a more general discussion on errors see \S6 in Bernardeau 
et al. 2002). We will consider two main sources of errors.  
One is due to the finite
number of particles ${\cal N}$ in the distribution. This error is usually
called the "Poisson error", and goes as: $\sim 1/{\cal N}$.  The second is due
to the finite size of the sample, which is characterized by:
\beq w_\Omega \equiv {1\over{\Omega^2}} \int \int
d\Omega_1 \Omega_2 ~w(\theta_{12})
\label{womega}
\eeq 
the mean correlation function over the solid angle of the survey $\Omega$.
This gives the uncertainty in the mean density on the scale of the sample,
which is constrained to be zero in most estimators, as the mean density is
calculated from the same sample, i.e. estimators suffer from the integral
constraint.  In general, this integral is not zero, but we need a clustering
model or a larger survey, such as the APM, to calculate its value.  For the
EDR size, $w_\Omega$ is dominated by the value of $w(\theta)$ on the scale of
the strip width: $w(\theta = 5 deg)$. From the APM $w(\theta)$ we find $\simeq
10^{-2}$ for EDR/S. For the APM size itself, 
this integral should be significantly
smaller, but its value is quite uncertain. For both for EDR and APM
sub-samples, the Poisson errors $\simeq 1/{\cal N} \simeq 10^{-5}-10^{-6}$ are
typically smaller than the sampling errors.  Neglecting Poisson errors and using
$w_N \sim q_N ~w^{N-1}$ for higher order correlations , Bernstein (1994) found:

\beq Cov(w_i,w_j) \simeq g(\gamma)~w_\Omega^2 + \beta~ w_\Omega
(w_i-w_\Omega) (w_j-w_\Omega)
\label{covar}
\eeq
where $g(\gamma)$ is a geometric term of order unity for power-law
correlations $w(\theta) \simeq A \theta^\gamma$. In the strict hierarchical
model, $w_N = q_N ~w^{N-1}$, we have $\beta =4(1-2 q_3+q_4)$. As this model 
is a rough approximation\footnote{It neglects the configuration and the scale
dependence of $q_3$ and $q_4$, which is only a good approximation
on non-linear scales, see Bernardeau et al. 2002}
 we will take $\beta$ to be a constant, which will be
fitted using the simulations.  The corresponding expression for the variance
(diagonal of the covariance) is:
\beq
Var(w_i) \simeq g(\gamma)~w_\Omega^2 + \beta~ w_\Omega
(w_i-w_\Omega)^2.
\label{var}
\eeq
In Fig.\ref{varw2} we compare the square root of the above expression $\Delta
w(\theta_i) \equiv \sqrt{Var(w_i)}$ with the RMS errors in $w_i$ from the
dispersion in 10 APM
sub-samples that simulate the geometry of the EDR/S sample.  
We find that a value of $\beta
\simeq 4$ fits well the above theoretical model to the errors in the
simulations. In principle, both $g$ and $\beta$ could be a function of scale,
but the model seems to match well the simulations, at least in the range
$\theta \simeq 0.1-4.0$ deg. On smaller scales we are approaching the map
pixel resolution and we should also include the variance due to the shot-noise
and finite cell-size. On scales larger that $\theta \simeq 4 deg$
we approach the EDR strip size and the integral constrain becomes important.
As we have not corrected for the integral constrain, we do not expect our
errors to follow the predictions on large scales. 
In the intermediate regime the model seems to work quite well.

Bernstein (1994) has shown, using Montecarlo simulations, that the model in
Eq.\ref{covar} works well for the covariance matrix. In his Fig.2 it shows the
covariance between adjacent bins $ Cov(w_i,w_{i+1})$. These predictions should
work well here if we compare alternative bins $ Cov(w_i,w_{i+2})$ instead
of adjacent bins, as we
are using 12 bins per decade as opposed to 6 bins per decade in Bernstein
(1994). The resulting covariance matrix is close to singular and most of its
principal components are degenerate. Thus, a significant test
estimation is not just straight forward.
 
With the help of Montecarlo simulations Bernstein (1994) concluded
that the effect of the off diagonal errors is small when fitting parametric
models, in particular a power-laws to $w(\theta)$. He find similar results for
the amplitude and the slope when using the simple diagonal
 chi-square minimization or the
of the principal components of the full covariance
matrix. Both the level of clustering and the errors in his Montecarlo 
simulations are quite similar to the ones presented here (compare left panel
of our Fig.\ref{varw2} to his Fig.1). Thus we can extend the conclusions 
of Bernstein (1994) to the  present analysis and, for simplicity, ignore the 
off-diagonal errors in the covariance matrix. In order to make sure that
the same conditions apply, we should use only every other bin 
in fitting models.

\subsection{An inflection point in $w_2(\theta)$?}

Right panel of Fig.\ref{varw2}
shows the logarithmic slope:
\beq
\gamma(r) = {d \log w_2(\theta) \over{d \log \theta}}
\eeq
of $ w_2(\theta)$ for the estimation in the second panel Fig.\ref{w2sdssNS}.
The mean and  errors in the top panel correspond to APM subsamples similar 
to the EDR/S. Within these
errors, both the APM and SDSS data are compatible with a power low
$w_2(\theta) \simeq \theta^\gamma$ with $\gamma$ between $\gamma \simeq -0.6$
and $\gamma \simeq -0.8$ (shown as two horizontal dotted lines), in good
agreement with Table 1 in Connolly etal (2001) and Maddox (1990).  Even with
this large errors there is a hint of a systematic flattening of $\gamma$
between 0.1 and 1.0 degrees in all subsamples.  This hint is clearer in the
combined analysis EDR/(N+S) where the errors (according to the APM subsamples)
are significantly smaller.  This flattening, of only $\Delta \gamma \simeq
0.1-0.2$ as we move from 0.1 to 1.0 degrees, it is apparent in all the APM and
SDSS subsamples.  It is reassuring that even at this detailed level all
data agree within the errors. It is also apparent from the top right panel of
Fig.\ref{varw2} that the errors are too large to detect this effect 
separately in EDR/S or EDR/N, so it depends on the good
calibration of EDR across the disjoint EDR/N and EDR/S samples.
 
The best fit to a power law model gives $\chi^2 \simeq 20$ for 10 degrees
of freedom, which corresponds to a $3\%$ confidence level for a power law
to be a good fit. If we do not use adjacent bins (see above \S 3.4) we
find  $\chi^2 \simeq 19$ for 5 degrees of freedom, which gives an even
lower confidence level.

\subsection{Smoothed 1-point Moments}

\begin{figure*}
\centerline{
\epsfxsize=8cm \epsfbox{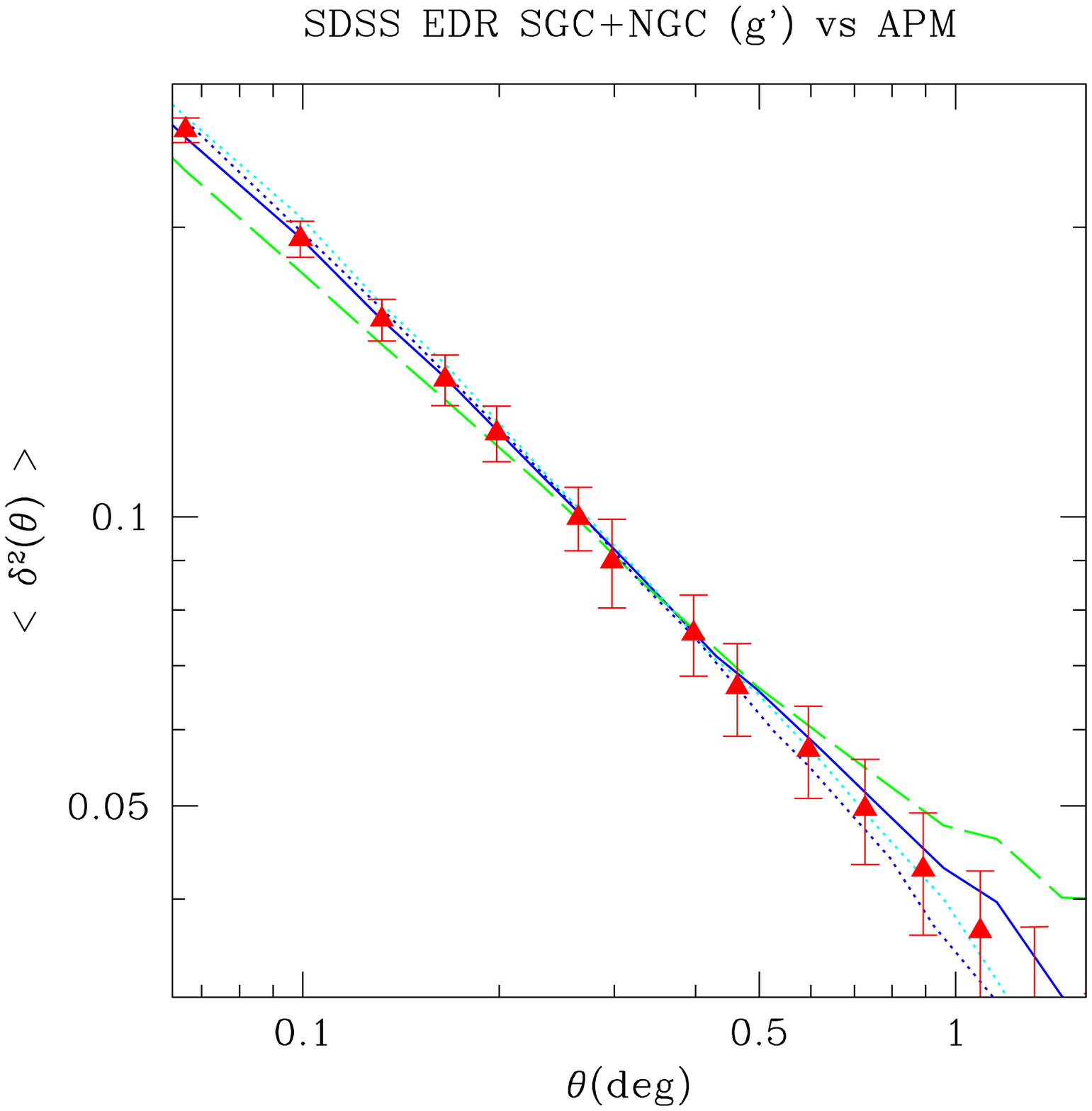}
{\epsfxsize=8cm \epsfbox{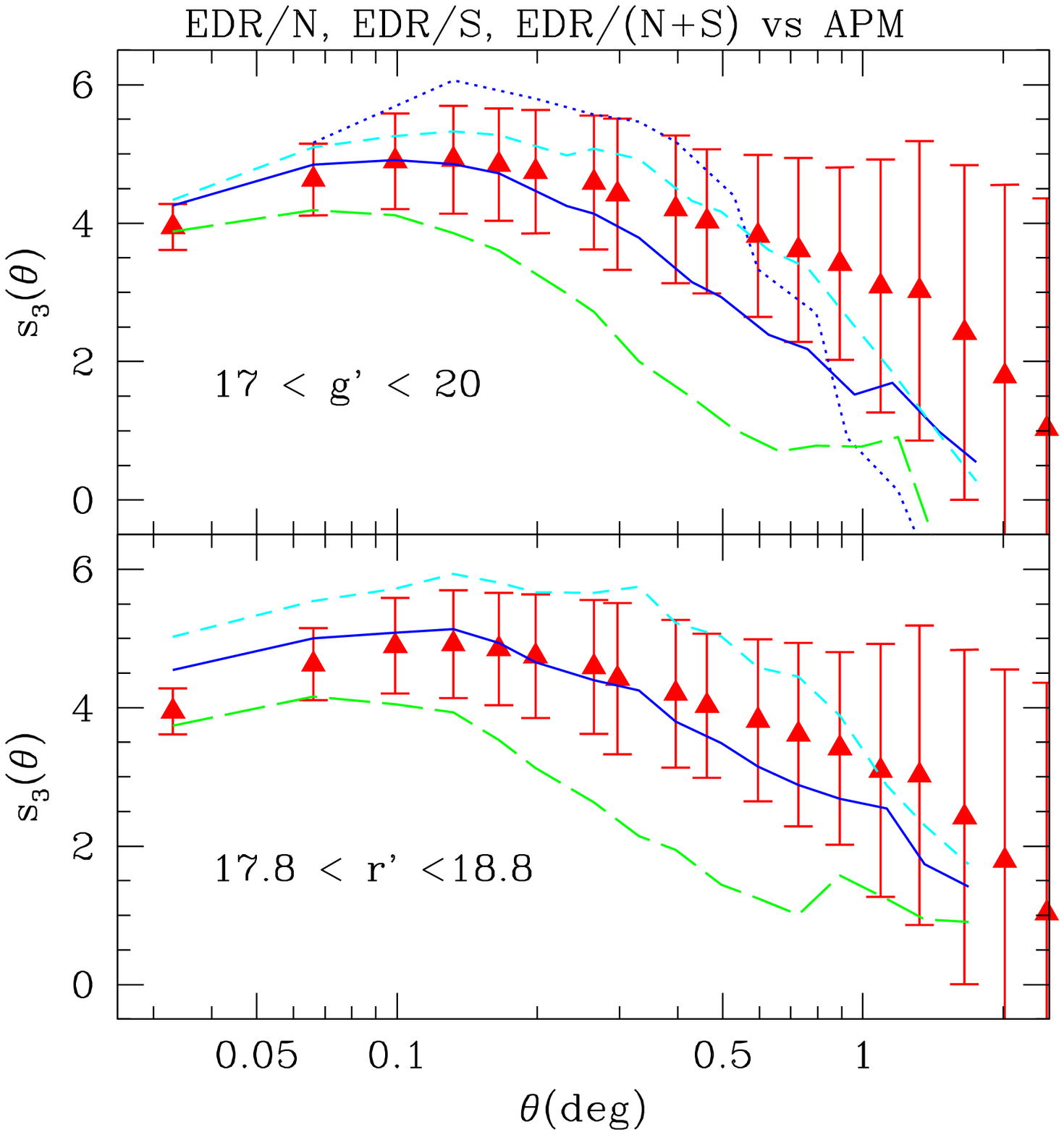}}
}
\caption[Fig2]{\label{X2sdssNSg} {\sc Left panel:} The variance  $\bar{w_2}$ 
as a function of angular smoothing $\theta$. Short and long dashed
lines correspond to the SDSS EDR/N and  EDR/S.
The dotted and continuous lines show EDR/(N+S) with and without the seeing mask.
The points with errorbars show 
the mean and 1-sigma confidence level in the values of
 10 APM sub-samples with same size and shape as the EDR/(N+S). \\
{\sc Right panel:} Same results for the reduced skewness $s_3$. 
The top and bottom panel show the
results in $g'$ and $r'$.}
\end{figure*}

We next compare the lower order moments of counts in cells of variable size
$\theta$ (larger than the pixel map resolution).  We follow closely the
analysis of Ga01.  Fig. \ref{X2sdssNSg} shows the variance of fluctuations in
density counts $\delta \equiv \rho/{\bar{\rho}} -1$ smoothed over a scale
$\theta$: $\bar{w_2} \equiv < \delta^2(\theta) >$, which is plotted as a
function of the smoothing radius $\theta$.  The errors show 1-sigma confidence
interval for APM subsamples with EDR/(N+S) size.  The individual results in
each subsample are strongly correlated so that the whole curve for each
subsample scales up and down within the errors, ie there is a strong covariance
at all separations due to large scale density fluctuations (eg see Hui \&
Gazta\~naga 1999 or Eq.\ref{womega} above).  
The EDR/(N+S) results (continuous and dotted lines) match perfectly well
the APM results, in agreement to what we found for the 2-point function above.
The size of the errorbars for EDR/N and EDR/S (not shown) are almost a factor
of two larger than for EDR/(N+S), so that they are also in agreement with the
APM within their respective sampling errors.

Right panel of
Fig. \ref{X2sdssNSg} shows the corresponding comparison for the normalized
angular skewness:
\beq
s_3(\theta) \equiv {<\delta^3(\theta)>\over{<\delta^2(\theta)>^2}} \equiv {\bar{w_2}(\theta)
\over{\bar{w_3}(\theta)}}
\eeq
All SDSS $g'$ sub-samples (top panel)
for $s_3$ show an excellent agreement with the APM at the
smaller scales (in contrast with the EDSGC results, see Szapudi \& Gazta\~naga
1998).  On larger scales the SDSS values are smaller, but the discrepancy is
not significant  given the strong covariance of individual APM subsamples.
Note how the effect of the seeing mask (dotted line) is to increase the
the amplitude of $S_3$, this could be partially due to systematic errors, but
it could also result from the smaller, $1/3$, sampling resulting from
 removing the pixels with bad seeing.

Bottom panel of the left of Fig. \ref{X2sdssNSg} shows the corresponding
results in $r'$.  At the smallest scale (of about 2' or 240 Kpc/h) we find
some slight discrepancies (at only the 1-sigma level for a single point) with the
APM.  The $r'$ results seem a scaled up version of the $g'$ results, which
indicates that the apparent differences could be explained in terms of
sampling effects (with strong covariance).  Note also that the value of $s_3$
seems to peak at slightly larger scale. This could indicate another explanation
for this discrepancy.  Szapudi \& Gazta\~naga 1998 argued that such peak could
be related to some systematic (or physical) effect related to the
de-blending of large galaxies.  It is reasonable to expect that such effect
could be strong function of color, as $g'$ and $r'$ trace different aspects of
the galaxy morphology.  We have also checked that results of individual scans
756 and 94 (and also 756+94) give slightly higher results, closer to the $r'$ results
than to the mean $g'$. Higher results are also found for the results with the
seeing mask (eg dotted line in top right panel of Fig. \ref{X2sdssNSg}).  
This goes in the right direction if we think that
de-blending gets worse with bad seeing, but it could also be affected by 
sampling fluctuations (because of the smaller area in the scans or masked data).

Similar results are found for higher order moments.  As we approach the scale
of $2$ degrees, the width of our strip, it becomes impossible to do counts for
larger cells and it is better to study the 3-point function.

\subsection{3-point Correlation function}

\begin{figure*} 
\centerline{
{\epsfxsize=8cm \epsfbox{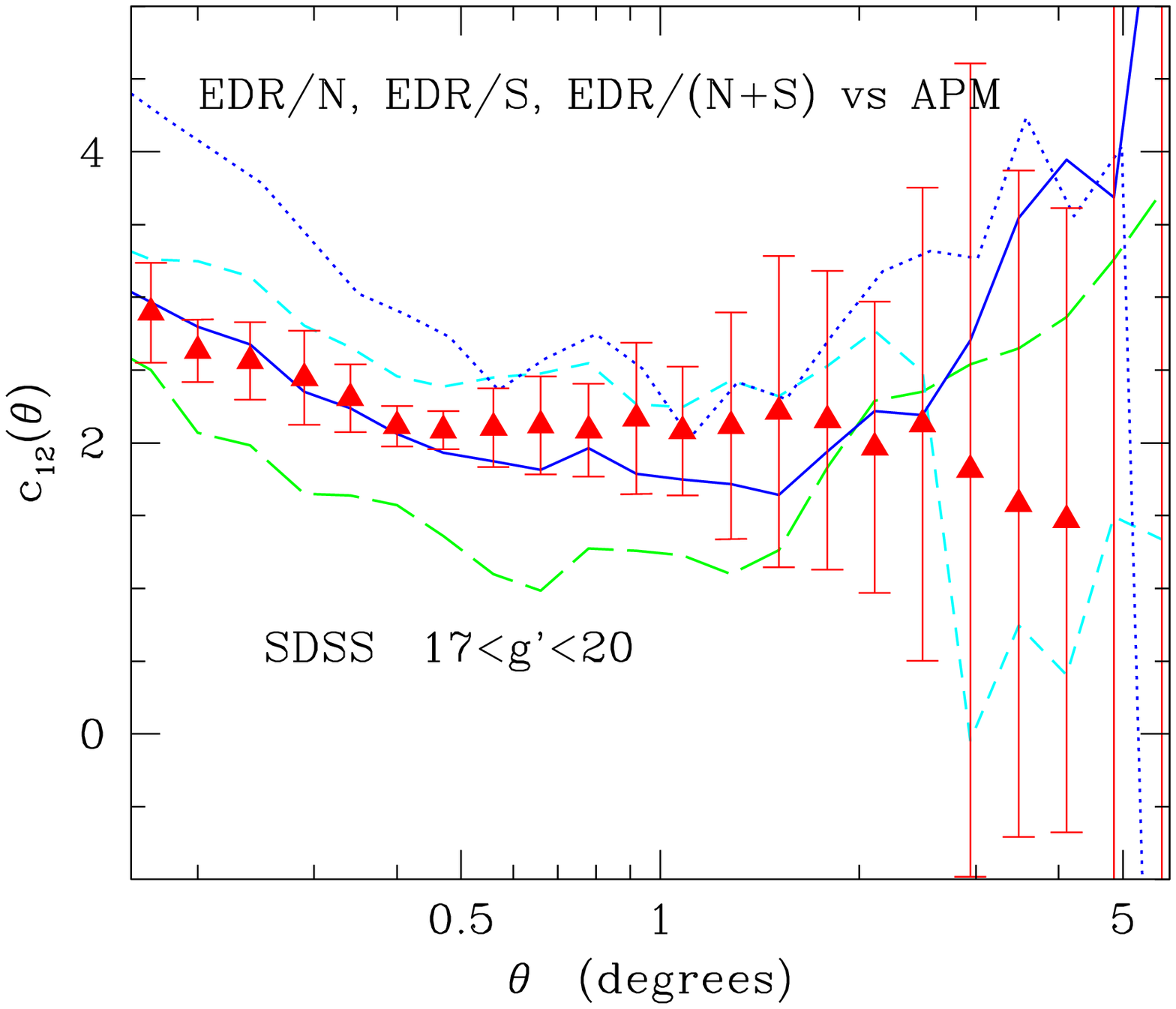}}
{\epsfxsize=8cm \epsfbox{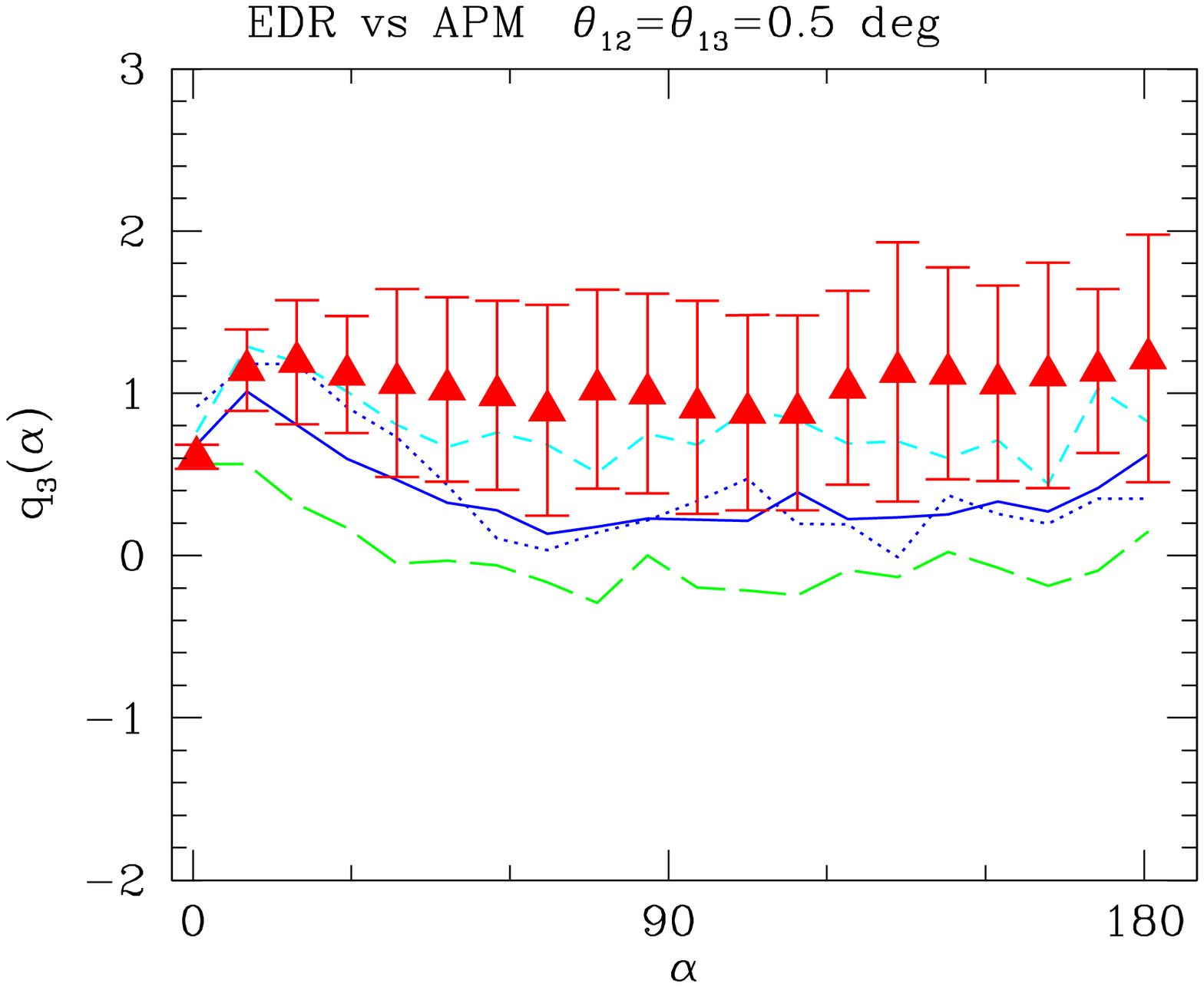}}
}
\caption[Fig2]{\label{c12sdssNS} 
{\sc Left panel:} The collapsed  3-point function  $c12$ as a function
of galaxy separation  $\theta$ for SDSS 
 $17<g'<20$. The short and long dashed
lines correspond to the SDSS EDR/N (North) and  EDR/S (South) strips. 
The dotted and continuous lines correspond to EDR/(N+S) with and without
the seeing mask shown in Fig.\ref{MapNSGCg20sei200}. The triangles with errorbars show
the mean and 1-sigma confidence level in the values of
10 APM sub-samples ($17<B_J<20$)
with same size as the joint EDR/(N+S) with masked seeing. 
\\
{\sc Right panel:}  Similar results for
the 3-point function $q_3(\alpha)$ in isosceles triangles of side
$\theta_{12}=\theta_{13}=0.5$ deg.}
\end{figure*}

Following Ga01 we next explore the 3-point function, 
normalized as:
\beq
q_3 \equiv {
w_3(\theta_{12}, \theta_{13}, \theta_{23}) \over
w_2(\theta_{12})w_2(\theta_{13})+w_2(\theta_{12})w_2(\theta_{23})+
w_2(\theta_{13})w_2(\theta_{23})}
\label{Qp}
\eeq
where $\theta_{12}$, $\theta_{13}$ and $\theta_{23}$ correspond to
the sides of the triangle form by the 3 angular positions
of $\delta_1 \delta_2 \delta_3$. Here we will
consider isosceles triangles, ie $\theta_{12}=\theta_{13}$,
so that $q_3=q_3(\alpha)$ is given as a function of the interior 
angle $\alpha$ which
determines the other side of the triangle $ \theta_{23}$ 
(Frieman \& Gazta\~naga 1999).

We also consider
the particular case of the collapsed configuration  $\theta_{23}=0$,
which corresponds to $ <\delta_1 \delta_2^2>$ and is normalized in slightly
different way (see also Szapudi \& Szalay 1999):
\beq
c_{12} \equiv {<\delta_1 \delta_2^2>\over{<\delta_1 \delta_2><\delta_1^2>}}
\simeq 2 q_3(\alpha=0) ~.
\eeq
Figure \ref{c12sdssNS} shows $c_{12}$ from the collapsed 3-point function.

Note the strong covariance in comparing the EDR/N to EDR/S.  The unmasked EDR/(N+S)
results agree well with the APM within errors, but the results with 
the seeing mask (shown as the dotted line) show significant departures at
small scales. 

Right panel in Fig. \ref{c12sdssNS} shows the reduced 3-point function
$q_3$ for isosceles triangles of side $\theta_{12}=\theta_{13}=0.5$ degrees.
Here there seems to be differences between APM and SDSS, but its
significance is low because the errors at a single point is only
1-sigma and that there is a strong covariance between points, eg note how the
EDR/S and EDR/N curves are shifted around the EDR(N+S) one. The APM seems closer
to the EDR/N.  This is a tendency that is apparent in previous
figures, but it is only on $q_3$ where the discrepancies starts to look
significant.  When estimation biases 
are present in the subsample mean, errors tend to be unrealisticly small, that is
errors are also biased down (see Hui \& Gazta\~naga 1999).

\section{Discussion and conclusions}

We have first explored the different uncertainties involved in the comparison
of the SDSS with the APM, such as the band or magnitude range to use.
After several test, we conclude that clustering in both the North and South EDR
strips (EDR/N and EDR/S) agree well in amplitude and shape with the APM on
scales $\theta < 2$ degrees. But we find inconsistencies with the APM
$w_2(\theta)$ at the level of $90\%$ significance on any individual scale at
$\theta > 2$ degrees.  This inconsistencies are larger than $90\%$ when we
compare EDR/S to EDR/N at any given point at $\theta > 2$ degrees
(compare the short and long dashed lines in left panel of Fig.\ref{w2sdssNSscan}).  
 We have shown that this is mostly
due to systematic photometric errors due to seeing variations
across the SDSS EDR (see right panel of Fig.\ref{w2sdssNSscan}).

We have pushed the comparison further by combining the North and South strips,
which we call EDR/(N+S) and analyze the EDR clustering as a whole.  Combining
samples in such a way is very risky  because small
systematic differences in the photometry tend to introduce large uncertainties
in the overall mean surface density. This was overcome in the APM by a
simultaneous match of many overlapping plates. For non-contiguous surveys the
task is almost impossible, unless one has very well calibrated photometric
observations, as is the case for the SDSS, to a level of 0.03 magnitudes (see
Lupton etal 2001).  The combined EDR/(N+S) sample shows very good agreement for
the number counts (see Figure \ref{ncountsall}) and also with the APM
$w_2(\theta)$, even at $\theta > 2$ degrees. In this case the agreement is in
fact within the corresponding sampling errors in the APM.  

Higher order correlations show similar results. The mean SDSS skewness is in
good agreement with the APM at all scales. The current SDSS sampling (1-sigma)
errors range from $10\%$ at scales of arc-minutes (less than 1 Mpc/h) to about
$50\%$ on degree scales ($\sim 10$ Mpc/h). At this level both surveys are in
perfect agreement. The collapsed 3-point function, $c_{12}$ shows even smaller
errors (this is because multi-point statistics are better sampled over narrow
strips than counts in large smoothed cells). At degree scales (which
correspond to the weakly non-linear regime $r\sim 8$ Mpc/h) 
we find $c_{12} \simeq 2.4 \pm
0.6$. This amplitude and also the shape is remarkably similar to that found
in simulations and what is theoretically expected from gravitational
instability $c_{12} \simeq 68/21 + 2/3\gamma$ (see Bernardeau 1996,
Gazta\~naga, Fosalba \& Croft 2001).  The 3-point function for isosceles
triangles of side $\theta_{12}=\theta_{13}=0.5$ deg.  (left panel in
Fig.\ref{c12sdssNS}) seems lower than the APM values, but within the 2-sigma
confidence level at any single point. Again here we would need of the
covariance matrix to say more.  In general, the North SDSS strip has higher
amplitudes for the reduced skewness or 3-point function than the Southern
strip.

We conclude that the SDSS is in good agreement with the previous galaxy
surveys, and thus with the idea that gravitational growth from Gaussian initial
conditions is most probably responsible for the hierarchical structures we see
in the sky (Bernardeau etal 2002, and references therein).

The above agreement has encourage us to look into the
detailed shape of $w_2(\theta)$ on intermediate scales, where the
uncertainties are smaller and errors from the APM are more reliable.  On
scales of 0.1 to 1  degree, we find indications of slight ($\simeq
20\%$) deviations from a simple power law (this is on larger scales than the
power law deviations found in Connolly etal 2001).  Right panel of Fig.\ref{varw2}
shows that the different SDSS samples have very similar slopes to the APM
survey, showing  a characteristic inflection with a
maximum slope.  In hierarchical clustering models, the initial slope of $\,
d\ln\xi/d\ln r$ is a smoothed decreasing function of the separation $r$.
Projection effects can partially wash out this curve, but can not produce any
inflection to the shape (at least if the selection function is also regular).
In Gazta\~naga \& Juszkiewicz (2001 and references therein) it was argued and
shown that weakly non-linear evolution produces a characteristic shape in
$\, d\ln\xi/d\ln r$. This shape, smoothed by projections, is evident in the
APM data for $\, d\ln w(\theta) /d\ln \theta$. Here we also find evidence
for such a shape in the combined EDR/(N+S) SDSS data.  The maximum in the
slope occurs around $\theta \simeq 0.6$ deg, which corresponds to $r \sim 5
$Mpc/h, as expected if biasing is small on those scales.

In summary, both the shape of the 2-point function and the shape and amplitude
of the 3-point function and skewness in the SDSS EDR data, confirms the idea
that galaxies are tracing the large scale matter distribution that started from
Gaussian initial conditions (Bernardeau etal 2002, and references therein).

\section*{Acknowledgments}

I would like to thank R.Scranton and J.Frieman for useful discussions
and comments on the first version of this paper.
I acknowledge support by grants from IEEC/CSIC and DGI/MCT BFM2000-0810 and
from coordinacion Astrofisica, INAOE.   Funding for the creation and distribution of the SDSS Archive has
been provided by the Alfred P. Sloan Foundation, the Participating
Institutions, the National Aeronautics and Space Administration, the National
Science Foundation, the U.S. Department of Energy, the Japanese
Monbukagakusho, and the Max Planck Society. The SDSS Web site is
http://www.sdss.org/.



\end{document}